\documentclass[12pt]{article}

\usepackage{amssymb}
\usepackage{amscd}
\usepackage{graphics}
\usepackage{amsfonts}
\usepackage{amsmath}
\usepackage{latexsym}
\usepackage[all]{xy}
\usepackage{fancyhdr}
\usepackage{color}
\usepackage{empheq}

\usepackage{latexsym,graphics,color}
\def\bes{\begin{subequations}}
\def\ees{\end{subequations}}
\def\ba{\begin{align}}
\def\ea{\end{align}}

\def\w{\wedge}

\def\be{\begin{equation}}
\def\ee{\end{equation}}

\def\br{\mathbb R}
\def\R{\mathcal R}

\def\k{\kappa}
\def\s1{\sigma^1}
\def\s2{\sigma^2}
\def\s3{\sigma^3}

\def\jp{\frac{1}{2}}
\def\ri{{\mathrm i}}

\definecolor{lila}{rgb}{1,0.2,0.9}
\definecolor{brown}{rgb}{0.5,0.3,0.3}
\definecolor{turquoise}{rgb}{0.2,0.9,0.7}
\definecolor{Orange}{rgb}{0.93,0.44,0}           
\definecolor{GrayBlue}{rgb}{0.35,0.4,0.62}       
\definecolor{SeafoamGreen}{rgb}{0.54,0.71,0.50}  

\definecolor{darkorange}{cmyk}{.20,.50,.80,0}
\definecolor{lightorange}{cmyk}{.07,.37,.65,0}
\definecolor{darkpeagreen}{cmyk}{.50,.30,.50,0}
\definecolor{lightpeagreen}{cmyk}{.22,.20,.40,0}

\def\ri{{\mathrm{i}}}                   %
\def\1{{\mbox{\boldmath $1$}}}          %
\def\ka{\kappa}
\def\lm{\lambda}                        %
\def\jp{\frac{1}{2}}                    %
\def\om{\omega}                         %
\def\Om{\Omega}                         %
\def\ga{\gamma}                         %

\definecolor{spec}{rgb}{0.0, 0.26, 0.15}
\def\bx{\boldsymbol{x}}
\def\bbp{\bar{\boldsymbol{p}}}
\def\bbx{\bar{\boldsymbol{x}}}
\def\bp{\boldsymbol{p}}

\def\bxi{\boldsymbol{\xi}}
\def\bbpi{\bar{\boldsymbol{\pi}}}
\def\bbxi{\bar{\boldsymbol{\xi}}}
\def\bpi{\boldsymbol{\pi}}

\def\bX{\boldsymbol{X}}
\def\bbP{\bar{\boldsymbol{P}}}
\def\bbX{\bar{\boldsymbol{X}}}
\def\bP{\boldsymbol{P}}
\def\bXi{\boldsymbol{\Xi}}
\def\bbPi{\bar{\boldsymbol{\Pi}}}
\def\bbXi{\bar{\boldsymbol{\Xi}}}
\def\bPi{\boldsymbol{\Pi}}

\def\bxi{\boldsymbol{\xi}}
\def\bpi{\boldsymbol{\pi}}

\def\bpm{\begin{pmatrix}}
\def\epm{\end{pmatrix}}

\DeclareMathSymbol{\Rho}{\mathalpha}{operators}{"50}

\begin{document}

\begin{flushright}
{}~
  
\end{flushright}

\vspace{1cm}
\begin{center}
{\large \bf  T-duality and T-folds for point particles}

\vspace{1cm}

{\small
{\bf Ctirad Klim\v{c}\'{\i}k}
\\
Aix Marseille Universit\'e, CNRS, Centrale Marseille\\ I2M, UMR 7373\\ 13453 Marseille, France}
\end{center}

\vspace{0.5 cm}

\centerline{\bf Abstract}
\vspace{0.5 cm}
\noindent  We argue that the T-duality phenomenon is not exclusively a stringy effect but it is relevant also in the context of the standard point particle dynamics. To illustrate the point, we construct a four-parametric family of four-dimensional
electro-gravitational backgrounds such that  the dynamics of a charged point particle in those backgrounds is insensitive   to a particular permutation of the parameters although this very  permutation does alter the background geometry.  
In particular, we find that a direct product of the Euclidean plane with   the two-dimensional Euclidean black hole  admits a point-particle T-dual with asymptotically  negative curvature. For 
 neutral particles, this point-particle  T-duality picture gets slightly modified because the T-duality map is no longer defined everywhere but only on a dense open domain of the  space of states. 
 We suggest a possible interpretation of this phenomenon in terms of a point particle T-fold.

  \vspace{2pc}
  
  \section{Introduction}
  
  T-duality in string theory is a phenomenon
  relating two geometrically inequivalent Kalb-Ramond-gravitational backgrounds via the dynamics of strings. Speaking more precisely, T-duality takes place when a   string moving in a background $(G,B)$ is dynamically equivalent to the string moving in the dual background $(\tilde G,\tilde B)$ even though the backgrounds $(G,B)$ and $(\tilde G,\tilde B)$ are not geometrically equivalent. Here by the "dynamical equivalence" is meant the existence of a canonical transformation transforming the Hamiltonian equations of motion of the string in the background $(G,B)$ 
  into  its Hamiltonian equations in the background $(\tilde G,\tilde B)$. In mathematical terminology, the T-duality between two inequivalent target geometries $(G,B)$ and $(\tilde G,\tilde B)$ is thus established if the
  phase space $P$ of the string moving in $(G,B)$ is symplectomorphic to the phase space $\tilde P$ of the string moving in $(\tilde G,\tilde B)$   and this
  symplectomorphism transforms the original Hamiltonian $H$ of the string into the dual one $\tilde H$.
  
  A point-particle analogue of the stringy Kalb-Ramond-gravitational background $(G,B)$ is naturally an electromagnetic-gravitational background $(G,A)$. Indeed, locally speaking, while
  string world-sheet couples naturally to the Kalb-Ramond two-form field $B$, the point-particle world-line couples naturally to the one-form field $A$ interpreted
  as the electromagnetic potential.  It is now evident what should be meant be the T-duality in the point-particle context. Indeed, we shall say that two geometrically inequivalent electromagnetic-gravitational backgrounds $(G,A)$ and $(\tilde G,\tilde A)$ are T-dual to each other if there exists a  canonical transformation transforming the point-particle Hamiltonian equations of  motion in the background $(G,A)$ 
  into  the Hamiltonian equations of motion in the background $(\tilde G,\tilde A)$. The main result of the present article is the  construction  of the four-parametric family of explicit  examples of such T-dual pairs of dynamically equivalent but geometrically inequivalent electromagnetic-gravitational backgrounds in four   dimensions.

  It appears that the point-particle T-duality has  not been so far suspected to exist. This is probably due to  the fact that there are no winding modes for point  particles which could be exchanged with the momentum modes. On the other
  hand, after the discovery of the Abelian T-duality with its momentum-winding exchange \cite{KY,SSe},
  more general T-dualities have been proposed in the framework of string theory  Ref. \cite{FJ,FT,DQ,KS95} for which  the question of the momentum winding exchange   is not  the central one (although it still can be posed, cf. Ref. \cite{KS97}). In fact, what  really matters for  the T-duality to take place is   the issue  of the existence of the canonical transformation relating the phase spaces associated to geometrically inequivalent backgrounds. With this shift of  perspective, the question of the existence of T-duality in the point particle dynamics should be treated without prejudices and  our result presented in this article
  should therefore look less surprising. 
 
It is perhaps worth mentioning    that our construction of the point-particle T-duality examples is the fruits of a combination of a conceptual approach and of educated guesses. Conceptually, we have taken a lot of inspiration
 by exploring the symplectic geometries of the
 so called Drinfeld doubles which are Lie groups  endowed with suitable symplectic forms. Such doubles have structures with certain (i.e. Poisson-Lie) dualities built in by construction, however, all point-particle T-duality examples which we constructed by considering various Drinfeld doubles suffered from the pathology that the dualizable Hamiltonian did not provide a complete flow on the domain of definition of the duality canonical transformation. Fortunately, we were able to cure some of those pathological examples "by hand",
 loosing of course the Drinfeld double interpretation but gaining the full-fledged,
  healthy and globally holding examples of
 the point-particle T-duality.
 
The plan of the article is as follows. In Section 2, we study in detail the basic building block of our   T-dualizable four-dimensional backgrounds which is certain two-parameter deformation of the Euclidean black hole in two dimensions \cite{Wi}.   In Section 3, we consider the direct product of two deformed black holes and we add to it a judiciously chosen background electric field. Then we describe the explicit canonical transformation relating this electro-gravitational background to its T-dual background, which is obtained by a suitable permutation of the four deformation parameters. We also  prove that the original and the dual backgrounds are  inequivalent as the Riemannian manifolds. In Section 4, we suppress the background electric field and we find that the T-duality symplectomorphism is then defined only on a dense open subset of the phase space. We speculate to interpret this phenomenon in terms of a point-particle analogue of the T-fold geometry known in string theory \cite{H}.  In Section 5, we provide conclusions and an outlook.

  \section{Deformed black hole geometry}
  
  Consider the standard polar coordinates $r,\phi$ on the
  plane $\br^2$ and a two-parametric family of Riemannian metrics
   \be  ds^2 = \frac{1}{1+\mu^2 r^{2}}dr^2  +   \frac{r^{2}}{1+\ga^2 r^{2}}  d\phi^2.\label{recast}\ee
  Whatever are the values of the parameters $\mu$ and $\ga$,  this metric exhibits no curvature singularity,  because  its Ricci scalar reads
 \be Ric=\frac{4\mu^2}{1+\ga^2 r^2}+\frac{6(\ga^2-\mu^2)}{(1+\ga^2 r^2)^2}.\label{rs}\ee
 The geometry \eqref{recast} is   asymptotically flat except for the case $\ga^2=0$, for which we obtain   the hyperbolic space of constant negative curvature.
 If $\ga^2=\mu^2=1$,  we recognize in the formula \eqref{recast}   the well-known Euclidean black-hole metric \cite{Wi}.  
 
  Consider an electric scalar  potential
  \be \varphi(r)=\jp\left(\ga^2+\frac{1}{r^2}\right).\label{ep}\ee
  The first order Hamiltonian dynamics of a charged point particle of a   charge $a^2$ in the geometry \eqref{recast} and in the electric potential \eqref{ep}  (no magnetic field!) takes then place in the
  phase space $P_a$, parametrized with the Darboux variables $p_r,r>0,p_\phi,\phi$. The Hamiltonian reads
  \be H_a=\jp g^{jk}(r)p_jp_k+\varphi(r)=\jp(1+\mu^2 r^2)p_r^2+\jp\left(\ga^2 +\frac{1}{r^2}\right)(p_\phi^2+a^2)\label{H2}\ee
  and   the symplectic form is
  \be \omega=p_r\w dr+p_\phi\w d\phi.\label{o}\ee
  In particular, the corresponding Hamiltonian equations of motion
  read 
 \be \dot p_r=-r\mu^2 p_r^2+\frac{p_\phi^2+a^2}{r^3},\quad \dot r=(1+\mu^2 r^2)p_r,\quad \dot\phi=\left(\ga^2 +\frac{1}{r^2}\right)p_\phi,\quad \dot p_\phi=0.\ee
A general solution of these equations depends on four real parameters $\zeta\in [0,2\pi[$, $\kappa >0$, $p_\phi$, $\nu$ and it reads
 
 \begin{subequations} \label{em}\begin{align} r(t)^2 &= \left(\frac{p_\phi^2+a^2}{\kappa^2}+\frac{1}{\mu^2}\right)\cosh^2{\left(\kappa\mu(t+\nu)\right)} -\frac{1}{\mu^2},\\
  p_r(t)r(t) &=\frac{\ka}{\mu}\tanh{(\ka\mu(t+\nu))},\\ \phi(t) &=\zeta+\ga^2 p_\phi t+\frac{p_\phi}{\sqrt{p_\phi^2+a^2}}
 \arctan{\left(\frac{\ka\tanh{(\ka\mu(t+\nu))} }{\mu \sqrt{p_\phi^2+a^2}}\right)}.\end{align}\end{subequations}
 We observe that for all possible values of the parameters $\zeta,\kappa,p_\phi,\nu$ the
 solution  \eqref{em} of the equation of motions is {\it complete}, which means that it avoids the  singularity $r=0$ and it does not arrive at infinity in a finite time $t$. 
 We thus conclude that the  Hamiltonian \eqref{H2} is non-pathological.

 \section{Charged particle}
 
 Consider a four-parametric four-dimensional background $T(\mu,\ga,m,c)$ obtained as 
 the direct product of two deformed black holes \eqref{recast} with the added electric potentials 
\begin{subequations}\label{times}\begin{align}  ds^2_\times &=\frac{1}{1+\mu^2 r^{2}}dr^2  +   \frac{r^{2}}{1+\ga^2 r^{2}}  d\phi^2+\frac{1}{1+m^2 \rho^{2}}d\rho^2  +   \frac{\rho^{2}}{1+c^2 \rho^{2}}  df^2,\label{ta}\\
\varphi_\times &=\jp\left(\ga^2+\frac{1}{r^2}\right)+\jp\left(c^2+\frac{1}{\rho^2}\right).\end{align}\end{subequations}
The dynamics of the charged point particle of the positive charge $a^2$ in the background $T(\mu,\ga,m,c)$ is then governed by the Hamiltonian
  \be H_{a\times}= \jp(1+\mu^2 r^2)p_r^2+\frac{1+\ga^2 r^2}{2r^2} (p_\phi^2+a^2)+\jp(1+m^2 \rho^2)p_\rho^2+\frac{1+c^2\rho^2}{2\rho^2} (p_f^2+a^2) \label{H2t}\ee
  and by the symplectic form
  \be \omega_\times=p_r\w dr+p_\phi\w d\phi+p_\rho\w d\rho+p_f\w df.\label{omt}\ee
 We now express the  coordinates $p_r,r>0,p_\rho,\rho>0, p_\phi,\phi,p_f,f$ on the phase space $P_{a\times}$  in terms of new coordinates $P_R,R>0,P_\R,\R>0, P_\Phi,\Phi,P_F,F$
 as follows
 \begin{subequations} \label{ct}\begin{align} r&=\R\frac{\sqrt{\R^2P^2_\R +P_\Phi^2+a^2}}{\sqrt{\R^2P^2_\R +P_F^2+a^2}},\quad p_r=P_\R\frac{\sqrt{\R^2P^2_\R +P_F^2+a^2}}{\sqrt{\R^2P^2_\R +P_\Phi^2+a^2}},\quad p_\phi=P_\Phi,   \\
 \rho&=R\frac{\sqrt{R^2P^2_R +P_F^2+a^2}}{\sqrt{R^2P^2_R +P_\Phi^2+a^2}}, \quad p_\rho=P_R\frac{\sqrt{R^2P^2_R +P_\Phi^2+a^2}}{\sqrt{R^2P^2_R +P_F^2+a^2}}, \quad 
p_f=P_F,\\
    f&=F+\frac{P_F}{\sqrt{P_F^2+a^2}}\arctan{\left(\frac{RP_R}{\sqrt{P_F^2+a^2}}\right)}-\frac{P_F}{\sqrt{P_F^2+a^2}}\arctan{\left(\frac{\R P_\R }{\sqrt{P_F^2+a^2}}\right)},\\
 \phi&=\Phi-\frac{P_\Phi}{\sqrt{P_\Phi^2+a^2}}\arctan{\left(\frac{R P_R}{\sqrt{P_\Phi^2+a^2}}\right)}+\frac{P_\Phi}{\sqrt{P_\Phi^2+a^2}}\arctan{\left(\frac{\R P_\R}{\sqrt{P_\Phi^2+a^2}}\right)}.\end{align}\end{subequations}
 The transformation \eqref{ct} is the diffeomorphism
 of the phase space $P_{a\times}$ with the inverse diffeomorphism given by
 \begin{subequations} \label{cti}\begin{align} R&=\rho\frac{\sqrt{\rho^2p^2_\rho +p_\phi^2+a^2}}{\sqrt{\rho^2p^2_\rho +p_f^2+a^2}},\quad P_R=p_\rho\frac{\sqrt{\rho^2p^2_\rho +p_f^2+a^2}}{\sqrt{\rho^2p^2_\rho +p_\phi^2+a^2}},\quad P_\Phi=p_\phi,   \\
 \R&=r\frac{\sqrt{r^2p^2_r +p_f^2+a^2}}{\sqrt{r^2p^2_r +p_\phi^2+a^2}}, \quad P_{\R}=p_r\frac{\sqrt{r^2p^2_r +p_\phi^2+a^2}}{\sqrt{r^2p^2_r +p_f^2+a^2}}, \quad 
P_F=p_f,\\
    F&=f-\frac{p_f}{\sqrt{p_f^2+a^2}}\arctan{\left(\frac{\rho p_\rho}{\sqrt{p_f^2+a^2}}\right)}+\frac{p_f}{\sqrt{p_f^2+a^2}}\arctan{\left(\frac{r p_r}{\sqrt{p_f^2+a^2}}\right)},\\
 \Phi&=\phi+\frac{p_\phi}{\sqrt{p_\phi^2+a^2}}\arctan{\left(\frac{\rho p_\rho}{\sqrt{p_\phi^2+a^2}}\right)}-\frac{p_\phi}{\sqrt{p_\phi^2+a^2}}\arctan{\left(\frac{r p_r}{\sqrt{p_\phi^2+a^2}}\right)}.\end{align}\end{subequations}

 Moreover,  the transformation \eqref{ct} is the symplectic diffeomorphism (or symplectomorphism) of the phase space $P_{a\times}$ because it preserves the symplectic form $\omega_\times$. Indeed, inserting the formulas \eqref{ct} into \eqref{omt} gives
  \be \omega_\times=dP_R\w dR+dP_\Phi\w d\Phi+dP_\R\w d\R+dP_F\w dF.\ee
It remains to show that the canonical transformation
\eqref{ct} can be interpreted as the T-duality symplectomorphism. For that, we express the Hamiltonian \eqref{H2t} in terms of the new Darboux coordinates $P_R,R>0,P_\R,\R>0, P_\Phi,\Phi,P_F,F$.
The result is  
  \be H_{a\times}= \jp(1+m^2 R^2)P_R^2+\frac{1+\ga^2R^2}{2R^2}(P_\Phi^2+a^2)+\jp(1+\mu^2 \R^2)P_\R^2+ \frac{1+c^2\R^2}{2\R^2}(P_F^2+a^2). \label{H2td}\ee
The  comparison of the formula \eqref{H2td} with  \eqref{H2t} shows that the role of the parameters
$m$ and $\mu$ got exchanged while the parameters
$c$ and $\ga$ remained in their places. Said in other words, the Hamiltonian \eqref{H2td} describes
the dynamics of the charged point particle in the
dual background $T(m,\ga,\mu, c)$
\begin{subequations}\label{dtimes}\begin{align}  \tilde ds^2_\times &=\frac{1}{1+m^2 R^{2}}dR^2  +   \frac{R^{2}}{1+\ga^2 R^{2}}  d\phi^2+\frac{1}{1+\mu^2 \R^{2}}d\R^2  +   \frac{\R^{2}}{1+c^2 \R^{2}}  dF^2,\\
\tilde\varphi_\times &=\jp\left(\ga^2+\frac{1}{R^2}\right)+\jp\left(c^2+\frac{1}{\R^2}\right).\end{align}\end{subequations}
To conclude the argument, that this point-particle T-duality  indeed does something non-trivial, it is sufficient to show that the
flipping of the parameters $\mu$ and $m$ may alter
the Riemannian geometry of the dual background $T(m,\ga,\mu, c)$ with respect to that of the original one
$T(\mu,\ga,m, c)$. For that, consider for example the background $T(0,0,1,1)$ with the metric  
\be  ds^2_\times = dr^2  +   r^2  d\phi^2+  \frac{d\rho^2+\rho^{2}df^2}{1+\rho^{2}}.\label{or}\ee
We see that this is the Riemannian geometry of the direct product of the Euclidean plane with the Euclidean black hole \cite{Wi}.
 The metric corresponding to the dual background
 $T(1,0,0,1)$ is
 \be \tilde ds^2_\times =\frac{1}{1+ r^{2}}dr^2  +  r^2 d\phi^2+ d\rho^2  +   \frac{\rho^{2}}{1+  \rho^{2}}  df^2.\label{du}\ee
 Using the formula \eqref{rs}, we find easily  
 the respective Ricci scalars of the metrics \eqref{or} and \eqref{du}
 \be Ric=\frac{4}{1+\rho^2},\quad \widetilde {Ric}=-2+\frac{6}{(1+\rho^2)^2}.\ee
 We thus observe  that the Riemannian geometries \eqref{or} and \eqref{du} are inequivalent, because
 $Ric$ is strictly positive while $\widetilde {Ric}$ acquires also negative values.

 \section{Neutral particle}

In this section, we shall discuss the dynamics of a
neutral particle in the background $T(\mu,\ga,m,c)$.
Since the charge $a^2$ vanishes, the electric potential plays no role and the background can be therefore considered as purely gravitational.  We start our analysis with
the two-dimensional metric \eqref{recast} corresponding to the deformed Euclidean black hole
  \be  ds^2 = \frac{1}{1+\mu^2 r^{2}}dr^2  +   \frac{r^{2}}{1+\ga^2 r^{2}}  d\phi^2.\label{recastbis}\ee
 The Hamiltonian of the neutral particle in the background
 \eqref{recastbis} reads
  \be H=\jp(1+\mu^2 r^2)p_r^2+\jp\left(\ga^2 +\frac{1}{r^2}\right)p_\phi^2\label{H20}\ee
  and  the symplectic form is as before
  \be \omega=p_r\w dr+p_\phi\w d\phi.\label{obis}\ee
  Now the neutral particle can reach the origin $r=0$ because there is no repulsive electrostatic potential which could prevent it. This means that the neutral phase space $P$ is slightly bigger
  then the charged one $P_a$ and the coordinate chart $p_r$, $r>0$, $p_\phi,\phi$ does not cover it all, so we prefer rather to work with  globally defined coordinates at the price that the rotational symmetry of the background will be less explicit. Thus, we introduce the metric
  on the plane $\br^2$ by the formula
\be ds^2=\frac{ 4 d\bx d\bar\bx+\ga^2(\bx d\bbx+\bbx d\bx)^2-\mu^2
(\bx d\bbx-\bbx d\bx)^2}{4(1+\ga^2\bx\bbx)(1+\mu^2\bx\bbx)}, \label{met}\ee
where
\be \bx=x^1+\ri x^2, \quad \bar\bx=x^1-\ri x^2\ee
 and $x^1,x^2$ are the standard global Cartesian coordinates on $\br^2$. Actually, the metric \eqref{met} is that \eqref{recast} of the deformed black hole, as it is easy to verify by setting
  \be \bx=re^{\ri\phi}.\label{tran}\ee
  As far as the corresponding neutral phase space $P$ is concerned,
  it can be globally described as $\br^4$  covered by the complex Darboux coordinates $\bx=x^1+\ri x^2$,  $\bp=p_1+\ri p_2$,
so that the symplectic form   reads
\be \om=\jp(d\bbp\w d\bx+d\bp\w d\bbx).\label{omo}\ee
 The standard Hamiltonian is found by inverting the metric \eqref{met}, which gives the formula
 \be H = \jp \bp\bbp +\frac{\mu^2}{8}(\bp\bbx+\bbp\bx)^2-\frac{\ga^2}{8} (\bp\bbx-\bbp\bx)^2.  \label{hgm} \ee
The Hamiltonian \eqref{hgm} and the symplectic form \eqref{omo} give rise to the Hamiltonian \eqref{H20}
  and
the symplectic form \eqref{obis}, upon the   canonical transformation
\be \bx=re^{\ri \phi},\quad \bp=\left(p_r+\frac{\ri p_\phi}{r}\right)e^{\ri \phi}.\label{tranp}\ee
The equations of motion of the neutral particle in the
coordinates $\bx,\bp$ then read
\be \dot\bx=\bp+\frac{\mu^2}{2}(\bp\bbx+\bbp\bx)\bx+\frac{\ga^2}{2}(\bp\bbx-\bx\bbp)\bx,\ee
\be \dot\bp=-\frac{\mu^2}{2}(\bp\bbx+\bbp\bx)\bp+\frac{\ga^2}{2}(\bp\bbx-\bx\bbp)\bp,\ee
  and their general solution, depending on four real parameters $\zeta\in[0,2\pi[$, $\k\geq 0$, $\nu$, $\lm$, turns out to be
\be \bx=e^{\ri(\zeta-\ga^2\lm\k t)}\left(\lm\cosh{(\mu\k(t+\nu))}-\ri \frac{\sinh{(\mu\k(t+\nu))}}{\mu}\right),\  \bp=\frac{-\ri\k e^{\ri(\zeta-\ga^2\lm\k t)}}{\cosh{(\mu\k(t+\nu))}}. \label{nsol}\ee
Similarly as in the charged case, the solutions \eqref{nsol} are complete for all possible values
of the parameters $\zeta$, $\ka$, $\nu$, $\lm$, therefore the neutral Hamiltonian \eqref{H20} is non-pathological. 

What is crucial for the discussion in the present section is the fact that the Hamiltonian \eqref{H20}
stays non-pathological even if we cut out the points from the neutral phase space $P$ for which the complex coordinate $\bp$ vanishes. Indeed, looking at Eqs. \eqref{nsol}, we observe that whenever a solution has a non-vanishing initial value  $\bp(t_0)$ it remains non-vanishing  for all times $t$. Thus we can define
a restricted phase space
\be P_r=\{(\bx,\bp)\in P;\ \bp\neq 0\},\label{res}\ee
with the symplectic form \eqref{omo}  and the Hamiltonian \eqref{H20} (both  understood as the restrictions of $\omega$ and $H$ to $P_r$).

Although the restricted dynamical system $(P_r,\omega_r,H_r)$ is   complete,
it lacks the geometrical interpretation because
it misses that static solutions \eqref{nsol} with $\ka=0$. However, it will inspire us to serve as a building block of an attempt to construct a point-particle T-fold.

\medskip

Consider now the four-dimensional purely gravitational background
$T(\mu,\ga,m,c)$ with the direct product metric
\eqref{met} now  written in global coordinates 
$\bx,\bxi$ of the    target space $\br^2\times \br^2$  
$$ ds^2=\frac{ 4 d\bx d\bar\bx+\ga^2(\bx d\bbx+\bbx d\bx)^2-\mu^2
(\bx d\bbx-\bbx d\bx)^2}{4(1+\ga^2\bx\bbx)(1+\mu^2\bx\bbx)}+$$\be +\frac{ 4 d\bxi d\bar\bxi+c^2(\bxi d\bbxi+\bbxi d\bxi)^2-m^2
(\bxi d\bbxi-\bbxi d\bxi)^2}{4(1+c^2\bxi\bbxi)(1+m^2\bxi\bbxi)}. \label{mettimes}\ee
The dynamics of the neutral point particle in the
background \eqref{mettimes} is now given by the "doubled"
Hamiltonian
\be H_\times=\jp \bp\bbp +\frac{\mu^2}{8}(\bp\bbx+\bbp\bx)^2-\frac{\ga^2}{8} (\bp\bbx-\bbp\bx)^2 +\jp \bpi\bbpi +\frac{m^2}{8}(\bpi\bbxi+\bbpi\bxi)^2-\frac{c^2}{8} (\bpi\bbxi-\bbpi\bxi)^2.\label{Htg}\ee
 and the doubled Darboux symplectic form
 \be \Om_\times=\jp(d\bbp\w d\bx+d\bp\w d\bbx)+\jp(d\bbpi\w d\bxi+d\bpi\w d\bbxi)\label{om}.\ee
The both quantities $H_\times$ and $\Om_\times$ are defined on the neutral direct product phase space $P_\times$ parametrized by the global Darboux coordinates $\bx,\bp,\bxi,\bpi$.
The 
solutions of the direct product dynamical system $(P_\times, H_\times,\Om_\times)$ are obviously obtained by doubling  Eqs.\eqref{nsol}, in particular, we see that they are complete and they remain complete on the following restricted phase space
\be   P_{r\times}=\{(\bx,\bp,\bxi,\bpi)\in P_\times;\ \bp\neq 0, \  \bpi\neq 0\},\label{restimes}\ee
with the symplectic form \eqref{om}  and the Hamiltonian \eqref{Htg} (both  understood as the restrictions of $\Om_\times$ and $H_\times$ to $P_{r\times}$).

Now we define a diffeomorphism of the restricted phase space $P_{r\times}$ by the formulas
\begin{subequations}\label{canp}\begin{align} \bX&=\frac{\bp}{2\vert \bpi\vert\vert \bp\vert}(\bbxi\bpi+\bxi\bbpi+\bx\bbp-\bbx\bp),\quad   \bP=\frac{\vert \bpi\vert}{\vert \bp\vert}\bp\\ \bXi&=\frac{\bpi}{2\vert \bpi\vert\vert \bp\vert}(\bbx\bp+\bx\bbp+\bxi\bbpi-\bbxi\bpi), \quad \bPi=\frac{\vert \bp\vert}{\vert \bpi\vert}\bpi.\end{align}\end{subequations}
Note that the diffeomorphism \eqref{canp} is involutive, which means that it is equal to its inverse, and it is also symplectic because it holds
 \be \Om_{r\times}=\jp(d\bbP\w d\bX+d\bP\w d\bbX)+\jp(d\bbPi\w d\bXi+d\bPi\w d\bbXi)\label{ombis}.\ee
 The restricted Hamiltonian in the capital variables now reads
{\footnotesize \be H_{r\times}=\jp \bP\bbP +\frac{m^2}{8}(\bP\bbX+\bbP\bX)^2-\frac{\ga^2}{8} (\bP\bbX-\bbP\bX)^2 +\jp \bPi\bbPi +\frac{\mu^2}{8}(\bPi\bbXi+\bbPi\bXi)^2-\frac{c^2}{8} (\bPi\bbXi-\bbPi\bXi)^2.\label{HTg}\ee}
 We observe that the restricted Hamiltonian $H_{r\times}$ expressed in the upper case
 variables has the same form as  in the lower case ones \eqref{Htg} except that the parameters $m$ and $\mu$ get  exchanged! This is of course similar as in the case of the charged particle but now the situation is not quite the same. Indeed, in the neutral case Eq. \eqref{canp} does not give a symplectomorphism of the whole phase
 space $P_{\times}$ but only of the restricted one $P_{r\times}$
 so the interpretation in terms of T-duality is not straightforward.
 
 \medskip
 
 We could say that in the neutral case the symplectomorphism 
 \eqref{canp} of the restricted phase space $P_r$ realizes a sort of local T-duality between the background  $T(\mu,\ga,m,c)$ (cf. Eq.\eqref{mettimes}) and the permuted one  $T(m,\ga,\mu,c)$, 
 but   the inspiration taken from string theory \cite{H} also suggests to  try to interpret \eqref{canp} as the gluing symplectomorphism defining  a point-particle T-fold.
 \medskip
 
 The  idea of the point particle T-fold construction
is not to restrict the phase $P_\times$ corresponding to the background  $T(\mu,\ga,m,c)$   in order to  relate it to the permuted background
  $T(m,\ga,\mu,c)$ but rather to extend it to a bigger phase space
  $P_{e\times}$   which would contain the unrestricted phase spaces of both
  original and dual backgrounds.  The extended phase space $P_{e\times}$ would be obtained by glueing 
 two identical charts $P_\times$ (the lower case one and the upper case one) by  the
  transition diffeomorphism   \eqref{canp}.
  Since each chart is the symplectic manifold and the transition diffeomorphism is the symplectomorphism, it seems that
  the space $P_{e\times}$ must be naturally symplectic manifold
and it could therefore  serve as the phase space of an extended dynamical system (i.e. the T-fold) containing at the same time the totality of the dynamics of the neutral  point particle in two geometrically different 
  backgrounds $T(\mu,\ga,m,c)$   and
  $T(m,\ga,\mu,c)$.
  
  In the case at hand, the suggested   construction of the T-fold fails, however. The reason is that the gluing
  symplectomorphism \eqref{canp} does not fulfil certain technical condition\footnote{This condition says, roughly speaking, that
the points lying "near" the boundary of the restricted phase space $P_{r\times}$ should  be mapped by the symplectomorphism \eqref{canp} to points lying "far" from the boundary.}  ensuring  that the resulting glued space be Hausdorff \cite{GQ}. We believe, however, that it is meaningful
  to continue  to look for another local T-duality symplectomorphism which would fulfil this technical condition  and would thus permit to construct a viable
  example of the point particle T-fold.

  \section{Conclusions and outlook}
  
The principal result of this article is the demonstration of the fact that the phenomenon of the T-duality exists not only in the stringy context but also in the point particle one.  The crucial role in the respect is played by the
explicit formulas \eqref{ct} and \eqref{canp} which realize the point particle
T-duality symplectomorphisms respectively for charged and neutral particle moving in the electro-gravitational background \eqref{times}. 

There are several open problems to be addressed in the context of the point-particle T-duality. Among others, it would be  nice to find out whether realistic  black hole solutions in four space-time dimensions admit point-particle T-duals or to clarify what is a physical relevance of the T-folds in the point-particle physics. However, arguably the most 
prominent open issue  is to work out the quantum status of the point-particle T-duality. Here the situation looks more promising than in string theory, where only the Abelian T-duality is under full control at the quantum level while all other known generalized T-dualities are so far  established only classically or were proven to take
place  up to one or two loops \cite{VKS,SST,CM,HR,BW}.


   \end{document}